\documentclass[prb,twocolumn,superscriptaddress]{revtex4}
\usepackage{graphicx}
\usepackage{amsmath}

\newcommand{\TF}{\protect\overrightarrow{t}}
\newcommand{\TB}{\protect\overleftarrow{t}}

\begin{document}
\title{Diode-like asymmetric transmission in ultrathin hyperbolic epsilon-near-zero slabs: \\ extreme anisotropy mimicking chirality}
\author{Carlo Rizza}
\affiliation{Department of Industrial and Information Engineering and Economics, Via G. Gronchi 18, University of L'Aquila, I-67100 L'Aquila, Italy}
\affiliation{Institute for superconductors, oxides and other innovative materials and devices, National Research Council (CNR-SPIN), Via Vetoio 10, I-67100 L'Aquila, Italy}
\author{Xin Li}
\affiliation{SUPA, School of Physics and Astronomy, University of St. Andrews, North Haugh, St Andrews, KY169SS, UK}
\author{Andrea Di Falco}
\affiliation{SUPA, School of Physics and Astronomy, University of St. Andrews, North Haugh, St Andrews, KY169SS, UK}
\author{Andrea Marini}
\affiliation{Department of Physical and Chemical Sciences, Via Vetoio 1, University of L'Aquila, I-67100 L'Aquila, Italy}
\author{Elia Palange}
\affiliation{Department of Industrial and Information Engineering and Economics, Via G. Gronchi 18, University of L'Aquila, I-67100 L'Aquila, Italy}
\author{Alessandro Ciattoni}
\affiliation{Institute for superconductors, oxides and other innovative materials and devices, National Research Council (CNR-SPIN), Via Vetoio 10, I-67100 L'Aquila, Italy}

\begin{abstract}
We demonstrate that a strong asymmetric transmission for forward and backward propagation of tilted circular polarized optical waves is supported by ultrathin epsilon-near-zero hyperbolic slabs. We find that, remarkably, this effect is solely triggered by anisotropy without resorting to any breaking of reciprocity and chiral symmetries or spatial nonlocal effects. In addition, we show that the asymmetric transmission undergoes a dramatic enhancement if the slab is hyperbolic in the epsilon-near-zero regime. This happens since, close to epsilon-near-zero point, the hyperbolic dispersion activates etalon resonances where extraordinary waves accumulate propagation phase even though the slab is ultrathin. The proposed strategy holds promise for realizing ultra-compact and efficient polarization devices at different frequency bands.
\end{abstract}


\maketitle

Media with very small permittivity exhibit a number of unique electromagnetic features providing a platform for the radiation manipulation at subwavelenght scales \cite{Liberal_1}. Epsilon-near-zero (ENZ) condition supports a static-like regime at a given frequency arising from the ``stretching" of the wavelength and this key ingredient has an enormous impact for nanophotonics applications shrinking, for example, the device size \cite{Zhong,Ciattoni_2}. Static-like features of the ENZ regime have been exploited to achieve an enormous variety of phenomena such as tunneling of electromagnetic waves through narrow channels \cite{Silveirinha}, developing highly directive emitters \cite{Alu}, slow-light \cite{Ciattoni_1,Newman} and geometry-invariant resonant cavities \cite{Liberal_2}.

\begin{figure}
\centering
\includegraphics[width=0.5\textwidth]{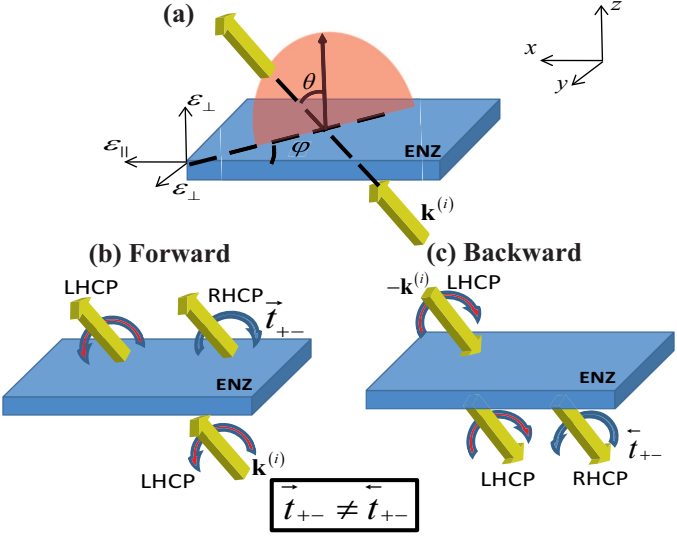}
\caption{Geometry of the asymmetric transmission process.
(a) Schematic view of the tilted plane wave (yellow arrow) impinging onto an ENZ hyperbolic slab (blue structure) and definition of the coordinate system and angular parameters. Propagation direction is defined as forward for the ${\bf k}^{(i)}$ direction (b) and backward for  
the $-{\bf k}^{(i)}$ direction (c). The setup supports circular asymmetric transmission: the forward cross-polarization transmission coefficient is not equal to the backward one. For example, we plot the situation where the incident wave is left-handed circular polarized (LHCP) and the transmitted wave has both the left-handed and the right-handed (RHCP) circular polarized component.}
\label{Figure1}
\end{figure}
%
%

%
%

In the context of nonlinear optics, ENZ media have attracted a great deal of interest since ENZ regime is able to trigger several non-linearity enhancement mechanisms. For example, the ubiquitous phase-matching, essential to achieve strong nonlinear coupling, is automatically satisfied in zero-index media \cite{Suchowski}. Furthermore, one can achieve high field intensity in a nonlinear ENZ medium due to the extreme effects appearing at the boundary where the normal electric field component undergoes a drammatic enhancement \cite{Campione,Vincenti,Ciattoni_4,Capretti,Luk}. In Ref.\cite{Ciattoni_3}, authors theoretically suggested that ENZ media enable the observation of extreme nonlinear effects since the nonlinear polarization is comparable with or even greater than the linear part of the overall dielectric response and, recently, such extreme nonlinear regime has been investigated in different setups \cite{Marini,Ciattoni_5,Ciattoni_6,Alam,Kaipurath,DiFalco,Prain}.
%
%

The strategy to reduce the predominant local contribution has also been exploited to emphasize the non-reciprocal and nonlocal responses \cite{Tretyakov,Davoyan}. Pollard \textit{et al.} have observed strong non-localities in ENZ metamaterials even supporting additional transverse magnetic waves \cite{Pollard}. Recently, Rizza \textit{et al.} \cite{Rizza} have shown that the electromagnetic response of a chiral ENZ  metamaterial is strongly affected by the first-order spatial dispersion.

An additional consequence of the ENZ static-like regime is the boosting of transverse magnetic (TM) and transverse electric (TE) asymmetric response in ultrathin ENZ slab. This effect has been demonstrated to support vortex generation \cite{Ciattoni_2}, efficient polarization conversion in ultrathin waveplates \cite{Ginzburg} and enhancement of the light spin-Hall effect \cite{Zhu}.


In this paper, we show that an ultrathin ENZ homogeneous slab can support a huge asymmetric transmission for forward and backward propagation of tilted circular polarized optical waves. Generally, asymmetric transmission and the associated diode-like response can be achieved by breaking either the Lorentz reciprocity or the spatial inversion symmetry. Reciprocity is broken in magneto-optical materials \cite{Bi,Wang}, in nonlinear media \cite{Lepri}, or in materials where the refractive index is spatio-temporal modulated \cite{Lira}, whereas spatial symmetry breaking is hosted by chiral materials and it has been investigated in several configurations \cite{Fedotov,Menzel,Pfeiffer,Fan}. In addition, Plum \textit{et al.} demonstrated that asymmetric transmission ensues from the mutual orientation of the structure and the incident light propagation direction (extrinsic chirality) \cite{Plum}. It is worth noting that such effect vanishes in the homogenized regime (i.e. when the structure period is much smaller than wavelength) or, in other words, it results from either the 1D chirality \cite{Rizza} or the second order spatial dispersion \cite{Gompf}. 

Here, we show that the unusual diode-like asymmetric transmission is merely triggered by the material anisotropic response since it entails, at oblique incidence, linear transmission dichroism and linear cross-polarization conversion. We prove that the diode-like response is dramatically enhanced if the slab is hyperbolic in the ENZ condition. This enhancement arises from the exotic feature of ENZ hyperbolic slabs of subwavelength thickness to exhibit etalon resonances \cite{Ciatt_optik}.


We consider a slab of thickness $L$ illuminated by a monochromatic plane wave of wavelength $\lambda$ whose electric field is ${\rm Re} \left[ {\bf E}(z) \exp{\left(-i 2 \pi c t/ \lambda +i k_y y +i k_x x \right) } \right]$ and impinging from vacuum with incident angles $\theta$ and $\varphi$, as reported in Fig.1(a), so that its wave vector is ${\bf k}^{(i)}={\bf k}_{\perp}+k_z^{(i)} {\hat{\bf e}}_z=k_0 \left( \sin \theta \cos \varphi {\hat{\bf e}}_x + \sin \theta \sin \varphi {\hat{\bf e}}_y +\cos \theta {\hat{\bf e}}_z \right)$  ($k_0=2\pi/\lambda$, ${\bf k}_{\perp}=k_x {\hat{\bf e}}_x+k_y {\hat{\bf e}}_y$). The considered slab is a homogeneous uniaxially anisotropic medium (a natural homogeneous material or a metamaterial), where the optical axis is oriented along the $x$-axis so that the electromagnetic response is described by the permittivity tensor $\varepsilon =\text{diag} (\varepsilon_{||}, \varepsilon_{\perp}, \varepsilon_{\perp})$. Imposing the transverse momentum conservation across the slab interfaces for a given transversal wave vector ${\bf k}_{\perp}$, there are two forward waves excited within the slab, i.e., the ordinary and extraordinary plane waves, whose their longitudinal wave vectors are
%
\begin{eqnarray}
\label{kk}
k_z^{(o)}&=& \sqrt{ k_0^2 \epsilon_{\perp}  - k_{\perp}^2}, \nonumber \\
k_z^{(e)}&=& \sqrt{k_0^2 \epsilon_{||}  -\left(k_x^2 \frac{\varepsilon_{||}}{\varepsilon_{\perp}}  +k_y^2 \right)}. \nonumber \\
\end{eqnarray}
%

In order to demonstrate the asymmetric transmission of circularly polarized waves as described in Fig.1 (b) and (c), we consider at first the the forward scattering in the TM (p) and TE (s) linear polarization basis $\hat{\bf e}_p = \cos \theta \cos\varphi {\hat{\bf e}}_x +\cos \theta \sin \varphi {\hat{\bf e}}_y - \sin \theta {\hat{\bf e}}_z$, $\hat{\bf e}_s = -\sin\varphi {\hat{\bf e}}_x +\cos \varphi {\hat{\bf e}}_y$. After matching the fields at the slab interfaces, we obtain the forward transmission matrix, $\overrightarrow{\bf T}_l$, connecting the components of the incident ${\bf E}^{(i)}$ and the transmitted ${\bf E}^{(t)}$ electric fields, namely
\begin{eqnarray}
 \begin{pmatrix}
   E_p^{(t)}  \\
   E_s^{(t)}  \\
  \end{pmatrix}
   =
  \begin{pmatrix}
   \TF_{pp} & \TF_{ps} \\
   \TF_{sp} & \TF_{ss} \\
   \end{pmatrix}
  \begin{pmatrix}
   E_p^{(i)}  \\
   E_s^{(i)}  \\
  \end{pmatrix}
  =\overrightarrow{\bf T}_l
  \begin{pmatrix}
   E_p^{(i)}  \\
   E_s^{(i)}  \\
  \end{pmatrix}.
\end{eqnarray}
Note that, for the considered slab, anisotropy yields to polarization conversion where the cross-polarization transmission coefficient are identical and non-vanishing, viz. $\TF_{ps}=\TF_{sp} \neq 0$ \cite{Li}.  On the other hand, in the circular polarization basis $\hat{\bf e}_+=1/\sqrt{2} \left( \hat{\bf e}_p+i \hat{\bf e}_s \right)$, $\hat{\bf e}_-=1/\sqrt{2} \left( \hat{\bf e}_p-i \hat{\bf e}_s \right)$, the forward transmission matrix
\begin{eqnarray}
&& \overrightarrow{\bf T}_c\
 =
  \left( {\begin{array}{cc}
   \TF_{++} & \TF_{+-} \\
   \TF_{-+} & \TF_{--} \\
  \end{array} } \right),
\end{eqnarray}
has the entries
\begin{eqnarray}
\label{circ-lin}
\TF_{++}&=&\frac{1}{2} \left(\TF_{pp}+\TF_{ss}\right), \nonumber \\
\TF_{+-}&=&\frac{1}{2} \left(\TF_{pp}-\TF_{ss}-i 2 \TF_{ps}\right),  \nonumber \\
\TF_{-+}&=&\frac{1}{2} \left(\TF_{pp}-\TF_{ss}+i 2 \TF_{ps} \right),  \nonumber \\
\TF_{--}&=&\frac{1}{2} \left(\TF_{pp}+\TF_{ss}\right),
\end{eqnarray}

Now, we consider the backward excitation as depicted in Fig.1(c) where the backward incident wave propagates along $-{\bf k}^{(i)}$.  
According to the reciprocal theorem \cite{Menzel_2}, asymmetric transmissions in the above linearly and circularly polarized bases are usually characterized by the parameters
\begin{eqnarray}
\Delta_l &=& |\TF_{sp}|^2-|\TB_{sp}|^2= |\TF_{sp}|^2-|\TF_{ps}|^2, \\
\Delta_c &=& | \TF_{+-} |^2 - |  \TB_{+-}|^2= | \TF_{+-} |^2 - |  \TF_{-+}|^2.
\end{eqnarray}
Using the above expressions for entries of the transmission matrices in the linear basis we obtain $\Delta_l=0$, i.e. asymmetric transmission does not occur for linearly polarized waves. Conversely, using Eqs.(\ref{circ-lin}), we get
\begin{equation}
\label{delta_c}
\Delta_c=2 |\Delta t| |\TF_{sp}| \sin \left( \Delta \psi \right),
\end{equation}
where we define $\Delta t= \TF_{pp}-\TF_{ss}$ and $\Delta \psi= Arg {\left( \Delta t   \TF_{ps}^* \right)}$.
From Eq.(\ref{delta_c}), it is evident that the asymmetric transmission for tilted plane waves is a consequence of
both the transverse electric and transverse magnetic polarization asymmetric response $\Delta t \neq 0$ and the presence of the linear cross-polarization conversion $\TF_{sp} \neq 0$, two ingredients naturally offered by an uni-axial medium.

%
%
\begin{figure}
\centering
\includegraphics[width=0.5\textwidth]{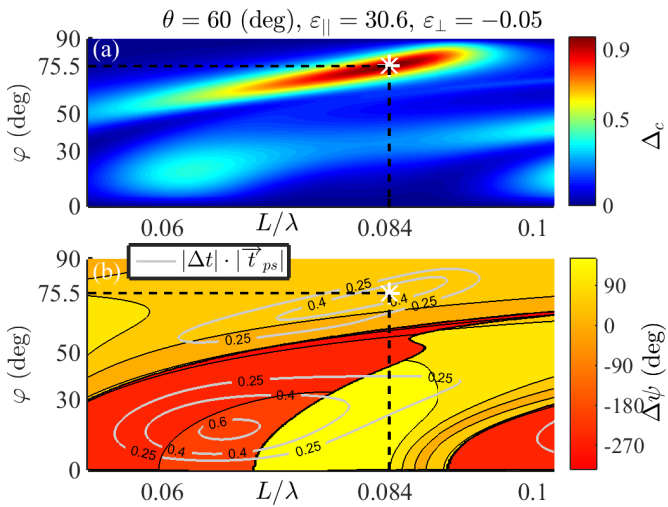}
\caption{(a) Asymmetric transmission $\Delta_c$ as a function of $\varphi$ and the normalized slab thickness $L/\lambda$.
(b) $\Delta \psi=Arg(\Delta t \TF_{ps}^*)$ as a function of $\varphi$ and the normalized slab thickness $L/\lambda$.  We obtain a near-unity asymmetric transmission for $|\Delta t| |\TF_{ps}|  \simeq 0.5$ and $\Delta \psi \simeq 90$ deg (marked with white stars). We contourplot the function $|\Delta t| |\TF_{ps}|$ with grey solid lines.} \label{Figure2}
\end{figure}

To demonstrate the enhancement of the asymmetric transmission in ENZ regime, we consider a slab made of monocrystalline bismuth
which is a homogeneous naturally-occurring material showing a hyperbolic response for far-infrared frequencies.
In this frequency range, the electromagnetic response of bismuth can be described by the Drude model,
\begin{eqnarray}
\label{BI}
\varepsilon_{\perp}&=&\epsilon_{\infty_{\perp}} \left( 1- \frac{\omega_{\perp}^2}{\omega^2 +i \omega \tau^{-1}}          \right), \nonumber \\
\varepsilon_{||}&=&\epsilon_{\infty_{||}} \left( 1- \frac{\omega_{||}^2}{\omega^2 +i \omega \tau^{-1}}          \right),
\end{eqnarray}
where $\omega=2 \pi c/\lambda$ is the angular frequency, $\epsilon_{\infty_{\perp}}=76$, $\epsilon_{\infty_{||}}=110$, $\tau=0.1$ ns is the relaxation time, $\omega_{\perp}=186$ cm$^{-1}$ and $\omega_{||}=158$ cm$^{-1}$ are the plasma frequencies \cite{Alekseyev}.
To discuss the role of ENZ condition, we neglect the material loss and we fix the angle $\theta=60$ deg and $\lambda=53.78$ $\mu$m (where $\varepsilon_{||}=30.6$ and $\varepsilon_{\perp}=-0.05$). In Fig.2(a), we plot the circular asymmetric transmission $\Delta_c$ as a function of the incident angle $\varphi$ and the normalized slab thickness $L/\lambda$. It is evident that the phenomenon of the asymmetric transmission is relevant for the considered configuration and it becomes maximum (viz. $\Delta_c=0.97$) at $\varphi=75.5$ deg and at $L=0.084 \lambda \simeq 4.5 $ $\mu$m (the white star in Fig.2(a)). As stated above, the effect is triggered from the transverse electric and transverse magnetic polarization asymmetric response and from the polarization conversion. In Fig.2(b), we highlight that these two phenomena are efficient and we obtain a near-unity asymmetric transmission for $|\Delta t| |\TF_{ps}|  \simeq 0.5$ (see grey contour lines) and $\Delta \psi \simeq 90$ deg (marked with the white star in Fig.2(b)). In the ENZ regime where $0 < \varepsilon_{\perp} \ll -1$ and $\varepsilon_{||} > 0$ and $k_{\perp} \neq 0$, we have that the ordinary wave is evanescent ($k_z^{(o)}= \sqrt{ - k_{\perp}^2}$), whereas the extraordinary wave is a propagating mode  and it can accumulate the desired propagation phase even if the slab is ultrathin. 

In fact, close to the ENZ condition, hyperbolic dispersion can support extraordinary waves whose longitudinal wave vectors are suitable large ($k_z^{(e)} \gg k_0$), thus activating even etalon resonances and enabling a huge diode-like response of the system. In summary, the effect is supported and triggered by the slab anisotropic response and it is dramatically enhanced in the ENZ hyperbolic regime.
%
%
\begin{figure}
\centering
\includegraphics[width=0.45\textwidth]{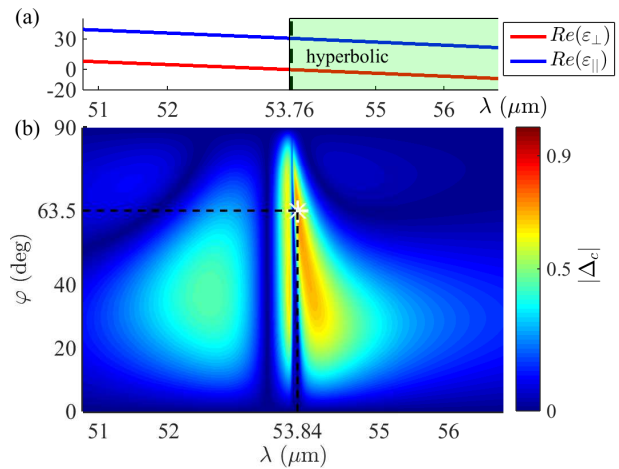}
\caption{(a) Real part of the bismuth dielectric constants $\varepsilon_{\perp}$, $\varepsilon_{||}$ reported in Eqs.(\ref{BI}) (the shadow area indicates hyperbolic region). The absolute value of $\Delta_c$ as a function of $\varphi$ and $\lambda$ for a bismuth slab with $L=4.5$ $\mu$m and $\theta=60$ deg. $|\Delta_c|$ attains its maximum ($\Delta_c=0.73$) for $\varphi=63.5$ (deg) and $\lambda=53.84$ $\mu$m (marked with a white star).} \label{Figure3}
\end{figure}
As expected, the imaginary part of $\varepsilon_{\perp}$, associated to the slab losses, plays a relevant role in the ENZ regime.
In Fig.3, we show the circular asymmetric transmission $|\Delta_c|$ as a function of $\lambda$ and $\varphi$ for a  bismuth slab, whose the dielectric response is given by Eqs.(\ref{BI}) and setting $L=4.5 $ $\mu$m and $\theta=60$ deg. The absolute value of $\Delta_c$ attains its maximum ($\Delta_c=0.73$) for $\varphi=63.5$ (deg) and $\lambda=53.84$ $\mu$m, where the slab presents an ENZ hyperbolic response (i.e., $\varepsilon_{||}=30.40+i0.02$ and $\varepsilon_{\perp}=-0.22+i0.02$).
Clearly, losses are not negligible in the ENZ regime, even if $\textrm{Im}(\varepsilon_{\perp}) \ll \textrm{Re}(\varepsilon_{\perp})$, so that their presence is detrimental and, in this case, we do not attain the ideal near-unity asymmetric transmission. Therefore, we conclude that the effect is modified by losses but, surprisingly, it persists in the situation where the slab is ultrathin.

In the considered example, we have discussed the asymmetric transmission occurring in a natural ENZ hyperbolic material.
Unfortunately, they are not available in some spectral region (e.g. in the optical range) \cite{Korzeb}, and thus one has to consider a metamaterial structure (whose response can be tuned by changing both its geometry and inclusion materials) for achieving the effect at the desired frequency.
In particular, we focus our attention on a subwavelength binary grating, made of silver and air whose filling fraction are $f_{Ag}=0.5$,$f_{Air}=0.5$, respectively, and whose stacking direction is along the $x$-axis.  The silver dielectric permittivity is described by the Drude model $\varepsilon_{Ag}=\epsilon_b-\omega_p^2/(\omega^2+i \alpha \omega)$ (where $\epsilon_b=5$, $\omega_p=14 \cdot 10^{15}$ Hz and $\alpha=0.32 \cdot 10^{15}$ Hz). Note that we increase the parameter $\alpha$ by a factor of $10$ to consider the layer size effect \cite{Cai}.
In the homogenized regime (namely, in the situation where $\Lambda \ll \lambda$ ), the effective electromagnetic response is described by the permittivity tensor $diag\left(\varepsilon_{||},\varepsilon_{\perp},\varepsilon_{\perp}\right)$, where $\varepsilon_{\perp}=f_{Ag} \varepsilon_{Ag} + f_{Air}$  and $\varepsilon_{||}=\left( f_{Ag}/ \varepsilon_{Ag} + 1/ f_{Air} \right)^{-1}$, so that the metamaterial response is characterized by the ENZ crossing point $\textrm{Re}(\varepsilon_{\perp})=0$ at $\lambda=0.4$ $\mu$m as reported in Fig.4(a). In Fig.4(b), we plot the $\Delta_c$ parameter evaluated for $\theta=60$ deg and $L=30$ nm. The asymmetric transmission attains its maximum $\Delta_c \simeq 0.2$ at $\lambda=0.42$ $\mu$m and $\varphi=45$ deg where $\varepsilon_{||} =1.32 + i 0.01$ and $\varepsilon_{\perp}= -0.14 + i 0.14$.

In order to verify the results of the effective medium theory (EMT) and to discuss the impact of spatial nonlocality on the considered asymmetric transmission, we consider the numerical implementation of the rigorous coupled-wave analysis technique \cite{Moharam} for obtaining the $0$th-order transmission coefficients and, hence, for evaluating the asymmetric transmission $\Delta_c$. In the Fig.4(c), for $\varphi=45$ deg and $\theta=60$ deg, we report the comparison between the asymmetric transmission evaluated with EMT (black solid line) and those predicted by the rigorous coupled-wave analysis for different values of the spatial period $\Lambda$. Clearly, close to the ENZ crossing point, non-locality affected the results \cite{Rizza} and, here, it slightly reduces the value of the $\Delta_c$ parameter. As a consequence, the results of the rigorous coupled-wave analysis technique assures that the asymmetric transmission is purely related to the structure homogeneous response.
\begin{figure}
\centering
\includegraphics[width=0.5\textwidth]{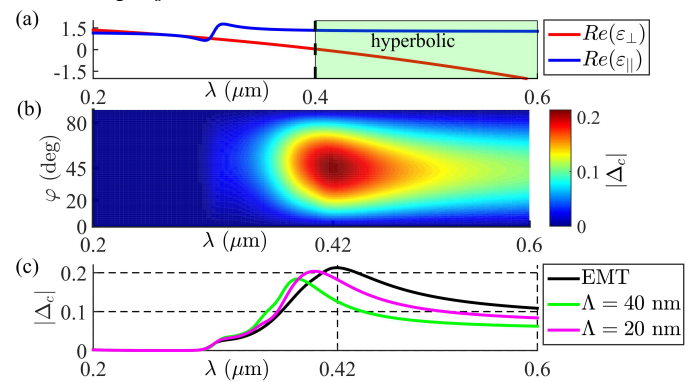}
\caption{(a) Real part of the metamaterial effective permittivity as a function of $\lambda$ (the shadow area indicates hyperbolic region). (b) $\Delta_c$ parameter as a function of $\varphi$ and $\lambda$ for $\theta=60$ deg and $L=30$ nm. (c) Comparison between the asymmetric transmission evaluated with EMT (black solid line) and those predicted by the rigorous coupled-wave analysis for different values of the spatial period $\Lambda$.} \label{Figure4}
\end{figure}

In conclusion, we demonstrate that an ultrathin ENZ hyperbolic slab supports a nearly 100\% asymmetric transmission for circularly polarized tilted waves. The optical like-diode response depends on neither non-reciprocity nor chirality nor spatial nonlocality and we show that it is merely triggered by the slab anisotropic response and it is dramatically enhanced in the ENZ hyperbolic regime. We believe that our investigation constitutes a fundamental step for the realization of compact polarization devices from THz to optical frequencies.

\section*{ACKNOWLEDGEMENT}
CR and AC thank the U.S. Army International Technology Center Atlantic for financial support (Grant No. W911NF-14-1-0315).
This work has partially been supported by the CNR-SPIN Seed Project No. B52F17001370005. AM acknowledges support from the ``Rita Levi Montalcini" programme for the recruitment of young researchers. CR and EP acknowledges support from the project NANOPREPAINT - PAR FSC Abruzzo 2007-2013 - Linea di Azione I.1.1.a.


\end{document}